# Perturbation of an Infinite Network of Identical Capacitors

**R. S. Hijjawi[†], J. H. Asad[††] ,A. J. Sakaji[†††] and J. M. Khalifeh[††]**

[†] *Department of Physics, Mutah University, Jordan*
[††]*Department of Physics, University of Jordan, Amman-11942, Jordan.*
[†††] *Department of Physics, Ajman University, UAE.*

## Abstract

The capacitance between any two arbitrary lattice sites in an infinite square lattice is studied when one bond is removed (i.e. perturbed). A connection is made between the capacitance and the Lattice Green's Function of the perturbed network, where they are expressed in terms of those of the perfect network. The asymptotic behavior of the perturbed capacitance is investigated as the separation between the two sites goes to infinity. Finally, numerical results are obtained along different directions and a comparison is carried out with the perfect capacitances.

## 1. Introduction

The calculation of the resistance of an infinite network of identical resistors is a classic well- studied problem in the electric circuit theory[1-3]. Despite this, the above mentioned problem is still of so interest and it brings the attention of many authors[4-6] to study and investigate it.

At the beginning of this century, this problem arises again in many publications[7,8]. The methods used in studying this problem vary from superposition of current distribution[4,5], random walk theory[6] and Lattice Green's Function (LGF) method[7,8].

In a recent work, the problem for many lattices for both the perfect and perturbed cases using Cserti's[7,8] method has been studied:

i- Asad[9] and Asad et. al[10] studied the perfect and perturbed (i.e. one bond is removed) square and simple cubic (SC) infinite lattices mathematically and experimentally. There was a good agreement between the mathematical and the experimental results especially for the bulk values. Also, there was a good agreement between our calculated values and those calculated by other previous authors[4-8].

ii- Asad et. al[11] studied the infinite perturbed (i.e. two bonds are removed) square lattice mathematically. Numerical results are obtained and a comparison with those obtained in Ref[9]. is carried out.

The LGF defined in our work is related to the Green's Function (GF) of the tight- binding Hamiltonian (TBH)[12]. The properties of the LGF for SC lattice have been studied in details especially when impurities are often introduced[13-17], where in these references some numerical results are presented and some recurrence formulae are used to calculate other values of the LGF. The analysis of the capacitance of an infinite network of identical capacitors has been arising recently[18-20]. Asad et. al[19] used the LGF method to calculate the capacitance between arbitrary lattice sites in a perfect infinite square lattice consisting of identical capacitors. While, Asad et. al. [20] used the supper position of charge distribution in calculating the capacitance between two points in an infinite square grid of identical capacitors. There were excellent agreements between the calculated results.

In this work, we shall use the Lattice Green's Function (LGF) approach to determine the capacitance for the so-called perturbed lattice obtained by removing one bond (capacitor) from the perfect infinite lattice. The content of this manuscript is helpful for electric circuit design and the method is instructive. As an example (see Fig. 1), consider an infinite square lattice consisting of identical capacitances C. Removing one bond from this perfect lattice results in a perturbed lattice.

## 2. Perfect Lattice

In this section, we reviewed the formalism of the perfect infinite network using Dirac's notation.

Consider a perfect d-dimensional infinite lattice consisting of identical capacitances $C$. All lattice points are specified by the position vector $\vec{r}$ given in the form

$$\vec{r} = l_1\vec{a}_1 + l_2\vec{a}_2 + ... + l_d\vec{a}_d .$$  (2.1)

where

$l_1, l_2, ..., l_d$ are integers (positive, negative or zero),

and

$\vec{a}_1, \vec{a}_2, ..., \vec{a}_d$ are independent primitive translation vectors.

Let the potential at the site $\vec{r}_i$ be $V(\vec{r}_i)$ and assuming a charge $Q$ enters the site $\vec{r}_i$ and a charge $-Q$ exits the site $\vec{r}_j$, while the charges are zero at all other lattice sites. Thus, one may write:

$$Q_m = Q[\delta_{mi} - \delta_{mj}] \quad \text{for all } m.$$  (2.2)

Then, according to Ohm's and Kirchhoff's laws we may write:

$$\frac{Q(\vec{r}_i)}{C} = \sum_{\vec{n}} [V(\vec{r}_i) - V(\vec{r}_i + \vec{n})].$$  (2.3)

where $\vec{n}$ are the vectors from site $\vec{r}$ to its nearest neighbors $(\vec{n} = \pm a_i, i = 1, 2, ..., d)$. One can form two state vectors, $V$ and $Q$ at the site $\vec{r}_i$ such that:

$$V = \sum_i |i\rangle V_i ;$$

$$Q = \sum_i |i\rangle Q_i .$$  (2.4)

where $V_i = V(\vec{r}_i)$ and $Q_i = Q(\vec{r}_i)$.

Here we assumed that $|i\rangle$, associated with the site $\vec{r}_i$, forms a complete orthonormal set, i.e. $\langle i|k\rangle = \delta_{ik}$ and $\sum_i |i\rangle\langle i| = 1$. Using Eq. (2.4) and Eq. (2.3) one gets:

$$\sum_j (z\delta_{ij} - \Delta_{ij})\langle j|V = \frac{\langle i|Q}{C}. \qquad (2.5)$$

$z$ is the number of neighbors of each lattice site (e.g. $z=2d$ for a d-dimensional hypercubic lattice).

and

$\Delta_{kl}$ is defined as:

$$\Delta_{kl} = \begin{cases} 1, & \vec{r}_k, \vec{r}_l \quad \text{are nearest neighbors} \\ \text{zero}, & \text{otherwise.} \end{cases} \qquad (2.6)$$

The summation in Eq. (2.5) is taken over all lattice sites. Multiplying Eq. (2.5) by $|i\rangle$ and summing over $i$, one gets:

$$\sum_{i,j} |i\rangle(\Delta_{ij} - z\delta_{ij})\langle j|V = \frac{-Q}{C}. \qquad (2.7)$$

Or one may write:

$$L_o V = \frac{-Q}{C}. \qquad (2.8)$$

Where $L_o = \sum_{i,j} |i\rangle(\Delta_{ij} - z\delta_{ij})\langle j|$.

$L_o$ is the so-called lattice Laplacian.

Similarly to the definition used in Economou[12], the LGF for an infinite perfect lattice can be defined as:

$$L_o G_o = -1. \qquad (2.9)$$

The solution of Eq. (2.9), which is a poisson-like equation, in its simple form, can be given as:

$$V = -\frac{L^{-1}_o Q}{C} = \frac{G_o Q}{C}. \qquad (2.10)$$

Inserting Eq. (2.2) into Eq. (2.10), we obtained:

$$V_k = \langle k | V = \frac{\langle k | G_o Q}{C};$$

$$= \frac{1}{C} \sum_m \langle k | G_o | m \rangle Q_m;$$

$$= \frac{Q}{C} [G_o(k,i) - G_o(k,j)]. \qquad (2.11)$$

where $\qquad G_o(l,m) = \langle l | G_o | m \rangle$ is the matrix element of the operator $G_o$ in the basis $|l\rangle$.

Finally, the capacitance between the sites $\vec{r}_i$ and $\vec{r}_j$ is then given as:

$$\frac{1}{C_o(i,j)} = \frac{V_i - V_j}{Q} = \frac{2}{C} [G_o(i,i) - G_o(i,j)]. \qquad (2.12)$$

The above formula can be rewritten as:

$$C_o(i,j) = \frac{C}{2[G_o(i,i) - G_o(i,j)]}. \qquad (2.13)$$

where we have used the fact that the LGF is symmetric (i.e. $G_o(l,m) = G_o(m,l)$).

It is interesting to study the asymptotic form of the capacitance for large separation between the sites $\vec{r}_i$ and $\vec{r}_j$. Using Eq. (2.30) in Ref. [19,20]. One can write

$$\frac{C_o(i,j)}{C} = \frac{1}{\frac{1}{\pi} (Ln\sqrt{(j_x - i_x)^2 + (j_y - i_y)^2} + \gamma + \frac{Ln8}{2})}. \qquad (2.14)$$

As the separation between the sites $\vec{r}_i$ and $\vec{r}_j$ goes to infinity then one finds that

$$\frac{C_o(i,j)}{C} \to 0. \qquad (2.15)$$

## 3. Perturbed Lattice

The charge contribution $\delta Q_i$ at the site $\vec{r}_i$ due to the bond $(i_o j_o)$ is given by

$$\frac{\delta Q_i}{C} = \delta_{i i_o}(V_{io} - V_{jo}) + \delta_{i j_o}(V_{jo} - V_{io}) \, ;$$

$$= \langle i | i_o \rangle (\langle i_o | - \langle j_o |) V + \langle i | j_o \rangle (\langle j_o | - \langle i_o |) V \, ;$$

$$= \langle i |(| i_o \rangle - | j_o \rangle)(\langle i_o | - \langle j_o |) V \, ;$$

$$= \langle i | L_1 V \, . \tag{3.1}$$

where the operator $L_1$ has the form

$$L_1 = (| i_o \rangle - | j_o \rangle)(\langle i_o | - \langle j_o |) \, . \tag{3.2}$$

Now, removing the bond $(i_o j_o)$ from the perfect lattice the charge $Q_i$ at the site $\vec{r}_i$ is given as:

$$(-L_o V)_i - \frac{1}{C} \delta Q_i = \frac{-Q_i}{C} \, . \tag{3.3}$$

Thus, Ohm's and Kirchhoff's laws for the perturbed lattice may be written as:

$$LV = \frac{-Q}{C} \, . \tag{3.4}$$

where

$$L_{o1} = L_o + L_1 \, . \tag{3.5}$$

Similarly to the perfect lattice, the LGF $G_{o1}$ for the perturbed lattice is defined as:

$$L_{o1} G_{o1} = -1 \, . \tag{3.6}$$

Therefore, Eq. (3.4) becomes;

$$V = \frac{G_{o1} Q}{C} \, . \tag{3.7}$$

Note that Eq. (3.7) is similar to Eq. (2.10). Here the operator $L_{o1}$ is now a sum of $L_o$ associated with the perfect lattice and a perturbation given by $L_1$.

To calculate the capacitance between the sites $\vec{r}_i$ and $\vec{r}_j$, we assume the charge to be given as in Eq. (2.2). Inserting Eq. (2.2) into Eq. (3.7) we get:

$$V_k = \langle k | V = \frac{\langle k | G_{o1} Q}{C} \, ;$$

$$= \frac{1}{C} \sum_m \langle k | G_{o1} | m \rangle Q_m \, ;$$

$$= \frac{Q}{C} [G_{o1}(k,i) - G_{o1}(k,j)]. \tag{3.8}$$

Thus, the capacitance between the lattice sites $\vec{r}_i$ and $\vec{r}_j$ is

$$\frac{1}{C_{o1}(i,j)} = \frac{V_i - V_j}{Q} \, ;$$

$$= \frac{1}{C} [G_{o1}(i,i) - G_{o1}(i,j) + G_{o1}(j,j) - G_{o1}(j,i)]. \tag{3.9}$$

The above formula can be rewritten as:

$$C_{o1}(i,j) = \frac{C}{[G_{o1}(i,i) - G_{o1}(i,j) + G_{o1}(j,j) - G_{o1}(j,i)]} \, . \tag{3.10}$$

Note here that $G_{o1}(i,i) \neq G_{o1}(j,j)$ since the translational symmetry is broken, but $G_{o1}(i,j) = G_{o1}(j,i)$. Our problem of finding the capacitances reduces to the calculation of the perturbed LGF, because once we calculate it then from Eq. (3.10) we find the perturbed capacitances on the networks.

Instead of the above idea, in the following we write the perturbed capacitance of the networks in terms of the perfect ones. Using Eqs. (2.5 and 2.6) we get:

$$(L_o + L_1)G_{o1} = -1. \tag{3.11a}$$

Or it can be rewritten as:

$$G_{o1} = -(L_o + L_1)^{-1}. \tag{3.11b}$$

Multiplying both sides of Eq. (3.11a) by $G_o$ we obtained the so-called Dyson's equation:

$$G_{o1} = G_o + G_o L_1 G_{o1} \, ;$$

$$= G_o + G_o L_1 G_o + G_o L_1 G_o L_1 G_o + \dots \tag{3.12}$$

To solve the above formula one can use the method presented by Economou[12]. Inserting Eq. (3.2) into the above equation, one gets:

$$G_{o1}(i,j) = \langle i | G_{o1} | j \rangle = G_o(i,j) + \frac{[G_o(i,i_o) - G_o(i,j_o)][G_o(i_o,j) - G_o(j_o,j)]}{1 - 2[G_o(i_o,i_o) - G_o(i_o,j_o)]}. \quad (3.13)$$

Finally, the capacitance between $\vec{r}_i$ and $\vec{r}_j$ can be obtained in terms of the perfect capacitances by using Eqs. (3.13, 3.10 and 2.13). Thus, we get after some simply straightforward algebra

$$\frac{C_{o1}(i,j)}{C} = \frac{1}{\dfrac{1}{C_o(i,j)} + \dfrac{[\dfrac{1}{C_o(i,j_o)} + \dfrac{1}{C_o(j,i_o)} - \dfrac{1}{C_o(i,i_o)} - \dfrac{1}{C_o(j,j_o)}]^2}{4[1 - \dfrac{1}{C_o(i_o,j_o)}]}}. \quad (3.14)$$

This is our final result for the perturbed capacitance between the sites $\vec{r}_i$ and $\vec{r}_j$ in which the bond $(i_o j_o)$ is removed.

To study the asymptotic behavior of the capacitance in an infinite perturbed square lattice for large separation between the sites $\vec{r}_i$ and $\vec{r}_j$, one can easily show from Eq. (3.14) and the fact that $C_o \to 0$ (see Ref. [19,20]) that also $C_{o1} \to 0$.

For symmetry reasons the capacitance between $\vec{r}_{i_o}$ and $\vec{r}_{j_o}$ in an infinite perfect d-dimensional lattice is $\dfrac{C_o(i_o,j_o)}{C} = d$, if $d \succ 1$. Then, using Eq. (3.14) the capacitance between the two ends of the removed bond is $\dfrac{C_{o1}(i_o,j_o)}{C} = d - 1$. For a square lattice ($d=2$) the capacitance is $C$ as mentioned in the introduction.

## 4. Numerical Results

In this section, numerical results are presented for an infinite square lattice including both the perfect and the perturbed case. The capacitance between the origin and the lattice site *(l, m)* in an infinite perfect square network has been calculated in Ref[18,19] (asad et. al).

On the perturbed square lattice the capacitance can be calculated from Eq. (3.14), and the calculated values of the perfect infinite square lattice mentioned above. In this work, the site $\vec{r}_i$ is fixed while the site $\vec{r}_j$ is moved along the line of the removed bond. Here we considered two cases; first, when the removed bond is between $i_o = (0,0)$ and $j_o = (1,0)$, where, our calculated values of the capacitances are arranged in Table 1 below. In the second case, the removed bond is shifted to be between $i_o = (1,0)$ and $j_o = (2,0)$, the calculated values of the capacitances are arranged in Table 2 below.

The capacitance between the sites of the removed bond is equal to $C$ as shown in Table 1 below. This capacitance can be considered as connected in parallel with a single capacitor of capacitance $C$, which gives a result of *2C* (the capacitance between two adjacent sites in an infinite perfect network of identical capacitance $C$).

In Figures 2-5 the capacitance is plotted against the site, for both the perfect and perturbed cases along [10] and [01] directions. Figs. 2 and 3 show the capacitance against the site for [10] direction. It is seen from those two figures that the perturbed capacitance is not symmetric along [10] due to the fact that the inversion symmetry of the lattice is broken along this direction. Also, as the broken bond is shifted away from the origin then the perturbed capacitance approaches the perfect one more rapidly. Figs. 4 and 5 the capacitance against the site for [01] direction. As shown from these two figures the inversion symmetry of the perturbed capacitance is not affected because there is no broken bond along this direction. From figures 4 and 5 one can see that as the broken bond along [10] direction is shifted away from the origin then the perturbed capacitance along [01] direction approaches the perfect one more rapidly.

Finally, one can see from figures 2-5 that for large separation between the sites $\vec{r}_i$ and $\vec{r}_j$, the perturbed capacitance goes to a finite value (i.e. $C_{o1} \to 0$) as mentioned before in section 3.

## Figure Captions

**Fig. 1** Perturbation of an infinite square lattice consisting of identical capacitances C by removing the bond between the sites $\vec{r}_{io}$ and $\vec{r}_{jo}$. The capacitance $C(i,j)$ is calculated between arbitrary lattice sites $\vec{r}_i$ and $\vec{r}_j$.

**Fig. 2** The capacitance on the perfect (squares) and the perturbed (circles) square infinite lattice between $i = (0,0)$ and $j = (j_x, 0)$ along the [10] direction as a function of $j_x$. The ends of the removed bond are $i_o = (0,0)$ and $j_o = (1,0)$.

**Fig. 3** The capacitance on the perfect (squares) and the perturbed (circles) square infinite lattice between $i = (0,0)$ and $j = (j_x, 0)$ along the [10] direction as a function of $j_x$. The ends of the removed bond are $i_o = (1,0)$ and $j_o = (2,0)$.

**Fig. 4** The capacitance on the perfect (squares) and the perturbed (circles) square infinite lattice between $i = (0,0)$ and $j = (0, j_y)$ along the [01] direction as a function of $j_y$. The ends of the removed bond are $i_o = (0,0)$ and $j_o = (1,0)$.

**Fig. 5** The capacitance on the perfect (squares) and the perturbed (circles) square infinite lattice between $i = (0,0)$ and $j = (0, j_y)$ along the [01] direction as a function of $j_y$. The ends of the removed bond are $i_o = (1,0)$ and $j_o = (2,0)$.

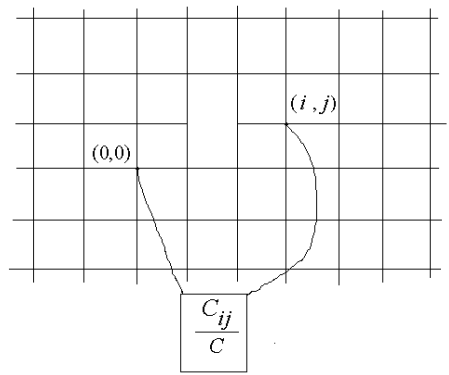

Fig. 1

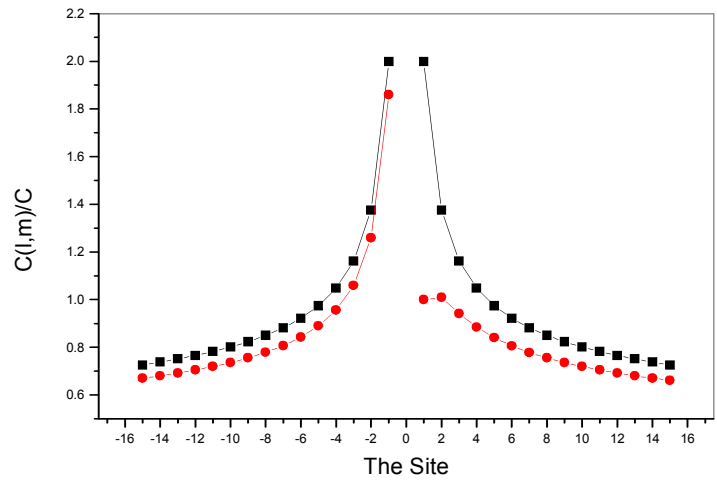

Fig. 2

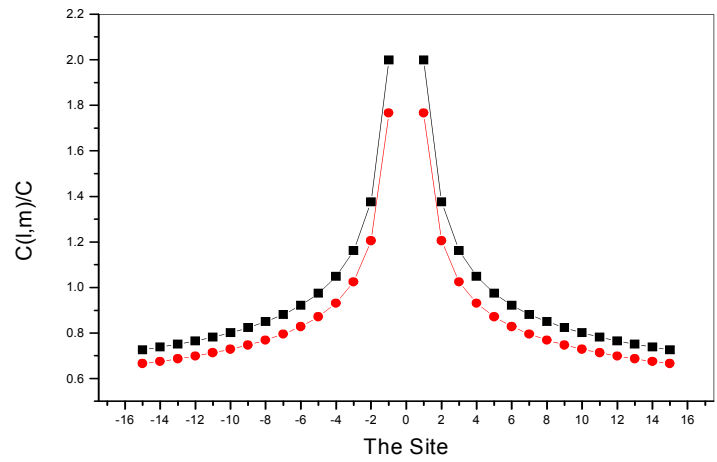

Fig. 3

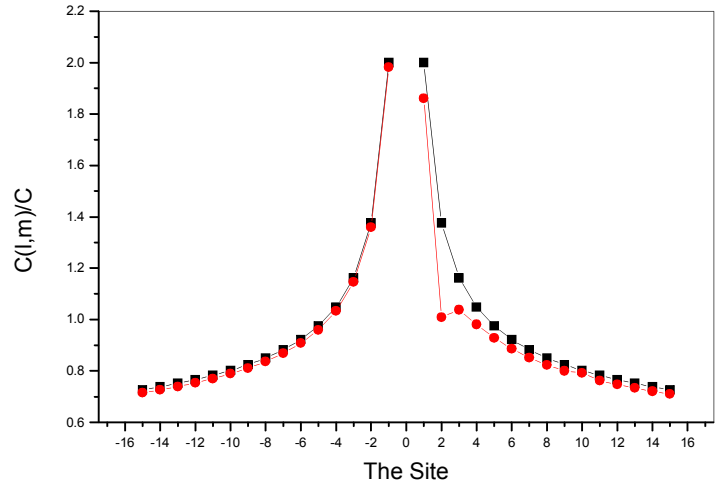

Fig. 4

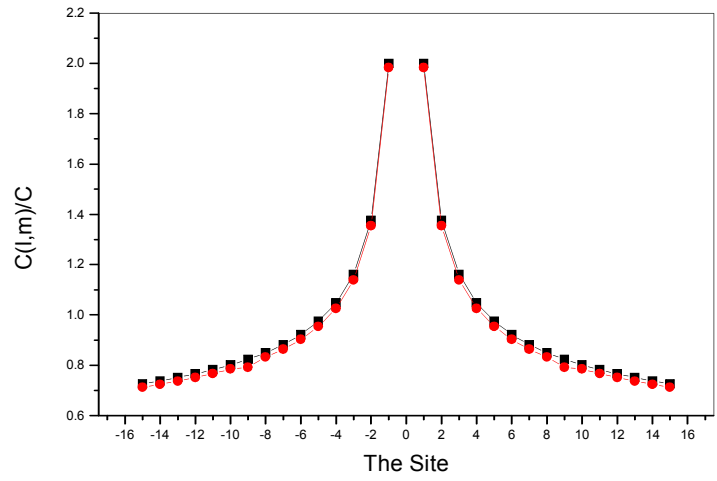

Fig.5

**Table Captions**

**Table 1:** The calculated capacitance between the origin and the site $j = (j_x, j_y)$, for an infinite square lattice in which the bond between the sites $i_o = (0,0)$ and $j_o = (1,0)$ is broken.

**Table 2:** The calculated capacitance between the origin and the site $j = (j_x, j_y)$, for an infinite square lattice in which the bond between the sites $i_o = (1,0)$ and $j_o = (2,0)$ is broken.

**Table 1:**

| $j = (j_x, j_y)$ | $\dfrac{C_{01}(i,j)}{C}$ | $j = (j_x, j_y)$ | $\dfrac{C_{01}(i,j)}{C}$ |
|---|---|---|---|
| (1,0) | 1 | (0,1) | 1.7667 |
| (2,0) | 1.0092 | (0,2) | 1.2054 |
| (3,0) | 0.9421 | (0,3) | 1.0249 |
| (4,0) | 0.8849 | (0,4) | 0.9314 |
| (5,0) | 0.8408 | (0,5) | 0.8714 |
| (6,0) | 0.8063 | (0,6) | 0.8285 |
| (7,0) | 0.7785 | (0,7) | 0.7956 |
| (8,0) | 0.7556 | (0,8) | 0.7692 |
| (9,0) | 0.7362 | (0,9) | 0.7475 |
| (10,0) | 0.7195 | (0,10) | 0.7291 |
| (11,0) | 0.7050 | (0,11) | 0.7132 |
| (12,0) | 0.6921 | (0,12) | 0.6993 |
| (13,0) | 0.6807 | (0,13) | 0.6870 |
| (14,0) | 0.6703 | (0,14) | 0.6760 |
| (15,0) | 0.6610 | (0,15) | 0.6661 |
| (0,0) | $\infty$ | (0,0) | $\infty$ |
| (-1,0) | 1.8610 | (0,-1) | 1.7667 |
| (-2,0) | 1.2597 | (0,-2) | 1.2054 |
| (-3,0) | 1.0601 | (0,-3) | 1.0249 |
| (-4,0) | 0.9563 | (0,-4) | 0.9314 |
| (-5,0) | 0.8902 | (0,-5) | 0.8714 |
| (-6,0) | 0.8433 | (0,-6) | 0.8285 |
| (-7,0) | 0.8076 | (0,-7) | 0.7956 |
| (-8,0) | 0.7793 | (0,-8) | 0.7692 |
| (-9,0) | 0.7561 | (0,-9) | 0.7475 |
| (1-0,0) | 0.7365 | (0,-10) | 0.7291 |
| (-11,0) | 0.7198 | (0,-11) | 0.7132 |
| (-12,0) | 0.7051 | (0,-12) | 0.6993 |
| (-13,0) | 0.6923 | (0,-13) | 0.6870 |
| (-14,0) | 0.6808 | (0,-14) | 0.6760 |
| (-15,0) | 0.6704 | (0,-15) | 0.6661 |

**Table 2:**

| $j = (j_x, j_y)$ | $\dfrac{C_{01}(i,j)}{C}$ | $j = (j_x, j_y)$ | $\dfrac{C_{01}(i,j)}{C}$ |
|---|---|---|---|
| (1,0) | 1.8610 | (0,1) | 1.9838 |
| (2,0) | 1.0092 | (0,2) | 1.3546 |
| (3,0) | 1.0379 | (0,3) | 1.1398 |
| (4,0) | 0.9813 | (0,4) | 1.0267 |
| (5,0) | 0.9284 | (0,5) | 0.9545 |
| (6,0) | 0.8858 | (0,6) | 0.9032 |
| (7,0) | 0.8516 | (0,7) | 0.8642 |
| (8,0) | 0.8235 | (0,8) | 0.8331 |
| (9,0) | 0.7999 | (0,9) | 0.7920 |
| (10,0) | 0.7923 | (0,10) | 0.7861 |
| (11,0) | 0.7624 | (0,11) | 0.7677 |
| (12,0) | 0.7470 | (0,12) | 0.7516 |
| (13,0) | 0.7334 | (0,13) | 0.7374 |
| (14,0) | 0.7212 | (0,14) | 0.7247 |
| (15,0) | 0.7102 | (0,15) | 0.7133 |
| (0,0) | $\infty$ | (0,0) | $\infty$ |
| (-1,0) | 1.9828 | (0,-1) | 1.9838 |
| (-2,0) | 1.3593 | (0,-2) | 1.3546 |
| (-3,0) | 1.1460 | (0,-3) | 1.1398 |
| (-4,0) | 1.0328 | (0,-4) | 1.0267 |
| (-5,0) | 0.9601 | (0,-5) | 0.9545 |
| (-6,0) | 0.9081 | (0,-6) | 0.9032 |
| (-7,0) | 0.8685 | (0,-7) | 0.8642 |
| (-8,0) | 0.8369 | (0,-8) | 0.8331 |
| (-9,0) | 0.8110 | (0,-9) | 0.7920 |
| (1-0,0) | 0.7891 | (0,-10) | 0.7861 |
| (-11,0) | 0.7704 | (0,-11) | 0.7677 |
| (-12,0) | 0.7534 | (0,-12) | 0.7516 |
| (-13,0) | 0.7396 | (0,-13) | 0.7374 |
| (-14,0) | 0.7268 | (0,-14) | 0.7247 |
| (-15,0) | 0.7152 | (0,-15) | 0.7133 |